\documentclass[10pt,twocolumn]{article}

\usepackage[a4paper,top=1.15cm,bottom=1.45cm,left=1.35cm,right=1.35cm,columnsep=0.58cm,headsep=0.30cm,footskip=0.82cm]{geometry}
\usepackage[T1]{fontenc}
\usepackage{lmodern}
\usepackage{amsmath,amssymb,bm}
\usepackage{graphicx}
\usepackage{booktabs}
\usepackage{xcolor}
\usepackage{mathtools}
\usepackage{array}
\usepackage{tabularx}
\usepackage{microtype}
\usepackage{titlesec}
\usepackage{caption}
\usepackage{fancyhdr}
\usepackage{dblfloatfix}
\usepackage{lineno}
\usepackage[numbers,square,sort&compress]{natbib}
\usepackage{hyperref}
\newcolumntype{Y}{>{\raggedright\arraybackslash}X}

\graphicspath{{./}{../}{figures/}{outputs/figures/}}

\definecolor{cpblue}{HTML}{0072B1}
\definecolor{cpgold}{HTML}{F2C300}
\hypersetup{hidelinks}

\titleformat{\section}{\sffamily\bfseries\fontsize{11.2}{12.5}\selectfont}{}{0pt}{}
\titlespacing*{\section}{0pt}{1.0em}{0.15em}
\titleformat{\subsection}{\sffamily\bfseries\fontsize{9.7}{10.8}\selectfont}{}{0pt}{}
\titlespacing*{\subsection}{0pt}{0.85em}{0.08em}
\titleformat{\subsubsection}{\sffamily\bfseries\fontsize{9.3}{10.3}\selectfont}{}{0pt}{}
\titlespacing*{\subsubsection}{0pt}{0.75em}{0.06em}

\DeclareCaptionLabelSeparator{cpbar}{\,|\,}
\newcommand{\safeincludegraphics}[2][]{%
    \IfFileExists{#2}{\includegraphics[#1]{#2}}{%
        \fbox{\parbox[c][4.1cm][c]{0.92\linewidth}{\centering
        Figure file not found:\\[2mm]\texttt{\detokenize{#2}}}}%
    }%
}

\newcommand{\ii}{\mathrm{i}}

\newcommand{\dd}{\mathrm{d}}

\fancypagestyle{firstpage}{%
  \fancyhf{}
  \fancyfoot[L]{\sffamily\scriptsize Communications Physics | Manuscript}
  \fancyfoot[R]{\sffamily\scriptsize\thepage}

}

\begin{document}

% Full-width first-page title/abstract block
\twocolumn[
\begin{@twocolumnfalse}
\noindent

%{\sffamily\bfseries\fontsize{20}{20}\selectfont communications}\,{\sffamily\fontsize{20}{20}\selectfont physics}
%\hfill {\sffamily\bfseries\fontsize{15}{16}\selectfont Article}\par
%\vspace{1mm}
%{\color{cpgold}\rule{\textwidth}{2.2pt}}\par
%\vspace{1mm}
%{\sffamily\footnotesize A Nature Portfolio journal}\par
%\vspace{10mm}
\hrule height 0.4pt
\vspace{2.2mm}

{\sffamily\bfseries\fontsize{24}{27}\selectfont
Complex mobility edges from competing localization channels in non-Hermitian quasicrystals \par}
\vspace{8mm}

\hrule height 0.4pt
\vspace{2.0mm}
{\sffamily\bfseries\fontsize{10.5}{12}\selectfont
Bing Shao\textsuperscript{1}, Weiwei Zhu\textsuperscript{2,*} \& Longwen Zhou\textsuperscript{1,3,4\textdagger}\par}
\vspace{3.4mm}
\hrule height 0.4pt
\vspace{3.0mm}

{\sffamily\fontsize{10.4}{13.1}\selectfont
{Non-Hermitian localization transitions often accompany multiple mobility edges, but their physical origin can be ambiguous. To address the issue, we introduce an extension of the {Ganeshan--Pixley--Das Sarma} model with both asymmetric hopping and complex onsite potential, thereby incorporating a broad range of non-Hermitian quasiperiodic chains. Its solution reveals two energy-dependent localization contributions manifested as distinct Lyapunov growth scales, whose competition determines the structure of mobility edges. We derive exact Lyapunov exponents via Avila's global theory, obtain thresholds of the two channels on complex-energy plane, and formulate a selection rule that distinguishes physical mobility edges from suppressed analytical candidates. The channel that uniquely controls localization can switch with energy, producing complex mobility edges of non-Hermitian origins. Nonreciprocal hopping further shifts the thresholds and alters the bulk energy patterns. Numerical spectra and finite-size localization diagnostics confirm the theory. Our work thus provides a unified framework for characterizing phase transitions and mobility edges under competing localization mechanisms in non-Hermitian quasicrystals.}\par}

\vspace{7mm}
\end{@twocolumnfalse}
]
\thispagestyle{firstpage}
\fontsize{9.0}{10.5}\selectfont

% Affiliation/correspondence note. Replace the placeholder before submission.
\footnotetext[1]{College of Physics and Optoelectronic Engineering, Ocean University of China, Qingdao 266100, China. \textsuperscript{\textdagger}zhoulw13@u.nus.edu.}
\footnotetext[2]{Center for Phononics and Thermal Energy Science, China-EU Joint Lab on Nanophononics, Shanghai Key Laboratory of Special Artificial Microstructure Materials and Technology, School of Physics Science and Engineering, Tongji University, Shanghai 200092, China. \textsuperscript{*}phyzhuw@tongji.edu.cn.}
\footnotetext[3]{Qingdao Key Laboratory of Advanced Optoelectronics, Qingdao 266100, China.}
\footnotetext[4]{Engineering Research Center of Advanced Marine Physical Instruments and Equipment of MOE, Qingdao 266100, China.}

% ============================================================
% I. INTRODUCTION
% ============================================================

\section{Introduction}

Quasicrystals exhibit long-range order without periodicity and have long served as ideal platforms for studying topological phenomena \cite{Kraus2012Pump,Kraus2012Equivalence,Verbin2013} and localization transitions \cite{Grempel1982,Sokoloff1985,Simon1985}. A paradigmatic example is the Aubry-Andr\'e-Harper (AAH) model \cite{Harper1955,Aubry1980,Jitomirskaya1999}, in which a localization transition occurs when the quasiperiodic potential exceeds a critical value corresponding to the self-dual point. Recent studies of various extensions of the AAH model have revealed a richer variety of transitions: introducing long-range hopping \cite{Biddle2010,Biddle2011,Deng2019}, generalized onsite potentials \cite{Soukoulis1982,Ganeshan2015,WangDuality2021,Cai2023,Banerjee2025}, or more complex lattice structures, such as mosaic and network configurations \cite{WangMosaic2020,Cui2026,HuHidden2025,HuSlow2025}, can induce energy-dependent localization~\cite{LiDasSarma2020,Lee2026Constraints}, allowing localized and extended states to coexist in different energy intervals, with the boundaries between them known as mobility edges. In addition, quasiperiodic localization and mobility edges have also been experimentally observed in quantum simulators such as ultracold atoms \cite{Roati2008,An2018Flux,Luschen2018,An2021} and photonic platforms \cite{Lahini2009,Weidemann2022,Lin2022Experiment,Chang2025Photonic,Lin2026Experiment}.

Non-Hermiticity~\cite{Ashida2020,Bergholtz2021,Bender2024}, introduced via asymmetric hopping or complex potentials, further enriches localization transitions and mobility-edge phenomena in quasicrystals~\cite{Jazaeri2001,HanZhou2022,Zheng2025}. For example, asymmetric hopping can drive delocalization under periodic boundary conditions~\cite{Hatano1996,Hatano1998,Jiang2019,Longhi2021AAH,ZhouGu2022}, whereas under open boundary conditions it can induce non-Hermitian skin effects~\cite{Kunst2018,YaoWang2018,Yokomizo2019,Okuma2020}, causing eigenstates to accumulate near the system boundaries~\cite{Liu2021a}. In addition, complex quasiperiodic potentials can induce {parity--time-reversal ($\mathcal{PT}$) symmetry} breaking~\cite{Longhi2019Topo,Liu2020PT,ZhouPT,ShaoPT} and produce elliptical complex-energy spectra~\cite{Longhi2019MIT} with mobility-edge loops~\cite{Longhi2021Maryland,Li2024Ring,Wang2024b,Pang2025,Chen2025NonAbelian}. Analytical results for mobility edges have been obtained for various non-Hermitian quasicrystals~\cite{Liu2020,ZengXu2020,Liu2021Dual,Liu2021c,Xia2022,Wang2025Multiple,He2026Network,Wang2026Maryland}. To date, however, the issue regarding how multiple mobility-edge loops coexisting in the complex-energy plane jointly determine the localization properties of the system has yet to be properly addressed.

Here, we reveal how competing localization channels determine complex mobility edges in a generic class of non-Hermitian quasiperiodic chain. Using Avila's global theory~\cite{Avila2015}, we derive the exact spectral Lyapunov exponent for a generalized non-Hermitian {Ganeshan--Pixley--Das Sarma (GPD)} model and identify two localization contributions associated with opposite hopping directions in momentum space. The quasiperiodic modulation makes these contributions depend differently on eigenenergy, allowing the channel-controlling localization to switch across the complex-energy spectrum.
This competition yields a selection rule for physical mobility edges. The stronger of the two localization contributions determines the actual transition boundary. Even when the spectrum crosses a candidate boundary, no localization transition occurs if the other, stronger channel dominates localization in that region. Physical mobility edges therefore consist of selected segments of the two candidate loops, with the dominant channel switching where these segments meet. Nonreciprocal hopping further shifts the localization thresholds and reshapes these boundaries. We verify this mechanism using numerical spectra and eigenstate localization diagnostics. These results show that the analytical loops jointly determine localization through channel competition, rather than each serving as an independent mobility edge.

% ============================================================
% II. MODEL
% ============================================================

\section{Results}
\label{sec:results}

% ============================================================
% II.1 MODEL AND THE ORIGIN OF ENERGY-DEPENDENT LOCALIZATION
% ============================================================
\subsection{The non-Hermitian GPD model}
\label{sec:model}

We start with a non-Hermitian extension of the GPD model, whose non-Hermiticity originates from both hopping asymmetry and complex onsite potentials. This model incorporates a broad range of non-Hermitian quasiperiodic chains as specific realizations while retaining analytically accessible localization thresholds. In real space, the Schr\"odinger equation of our model reads
\begin{equation}
    t e^{g}\psi_{n+1} + t e^{-g}\psi_{n-1} + V(\theta_n)\psi_n = E\psi_n,
    \label{eq:model}
\end{equation}
where
\begin{equation}
    V(\theta) = \frac{ 2\lambda\cos\theta + 2\ii\gamma\sin\theta }{ 1-a\cos\theta-\ii b\sin\theta }, \qquad \theta_n=2\pi\alpha n+\varphi .
    \label{eq:potential}
\end{equation}
The symmetric part of the hopping amplitude $t>0$ fixes the energy scale. $g\in\mathbb{R}$ controls the hopping asymmetry. The lattice index $n\in\mathbb{Z}$, and the potential parameters $\lambda$, $\gamma$, $a$ and $b$ are all real. Quasiperiodicity is realized by an irrational $\alpha$, set as $(\sqrt{5}-1)/2$ and approximated by Fibonacci ratios in numerical calculations. The parameter $\varphi\in[0,2\pi)$ represents a phase shift. We work in the pole-free regime $a^2+b^2<1$, which is sufficient to keep the denominator of $V(\theta)$ nonzero for any $\theta\in{\mathbb R}$. Figure~\ref{fig:model}(a) illustrates the real-space chain. All numerical spectra and right-eigenstate diagnostics in this work use periodic boundary conditions (PBC).
In the Hermitian GPD model, the denominator $1-a\cos\theta$ of $V(\theta)$ makes the effective momentum-space hopping energy dependent. States at different energies therefore reach the localization threshold at different potential strengths, allowing localized and extended states to coexist in the spectrum and be separated by a mobility edge~\cite{Ganeshan2015}. This mobility edge is found to be greatly enriched by non-Hermitian effects in our generalized GPD model, as will be uncovered below.

% ============================================================
% II.2 DUAL HOPPING AND COMPETING LOCALIZATION CONTRIBUTIONS
% ============================================================
\subsection{Unequal dual hopping and competing Fourier channels}
\label{sec:channels}

We now examine how the non-Hermitian modulation changes this energy-dependent localization mechanism. We first set $g=0$, and restore nonreciprocal hopping in the final subsection. To simplify the notation, we introduce $z=e^{\ii\theta}$, let $A=(a+b)/2$, $B=(a-b)/2$, and $c_\pm=\lambda\pm\gamma$. We can then write the onsite potential as $V(z)=v(z)/d(z)$ in Eq.~(\ref{eq:potential}), where $v(z)=c_+z+c_-z^{-1}$ and $d(z)=1-Az-Bz^{-1}$. In particular, $Ed(z)-v(z)=E-p_+(E)z-p_-(E)z^{-1}$, where $p_+(E)=AE+c_+$ and $p_-(E)=BE+c_-$. Thus, the denominator of $V(z)$ introduces $E$ into both coefficients $p_\pm$ before any further transformations are performed.

\begin{figure}[!t]
    \centering
    \safeincludegraphics[width=0.95\linewidth]{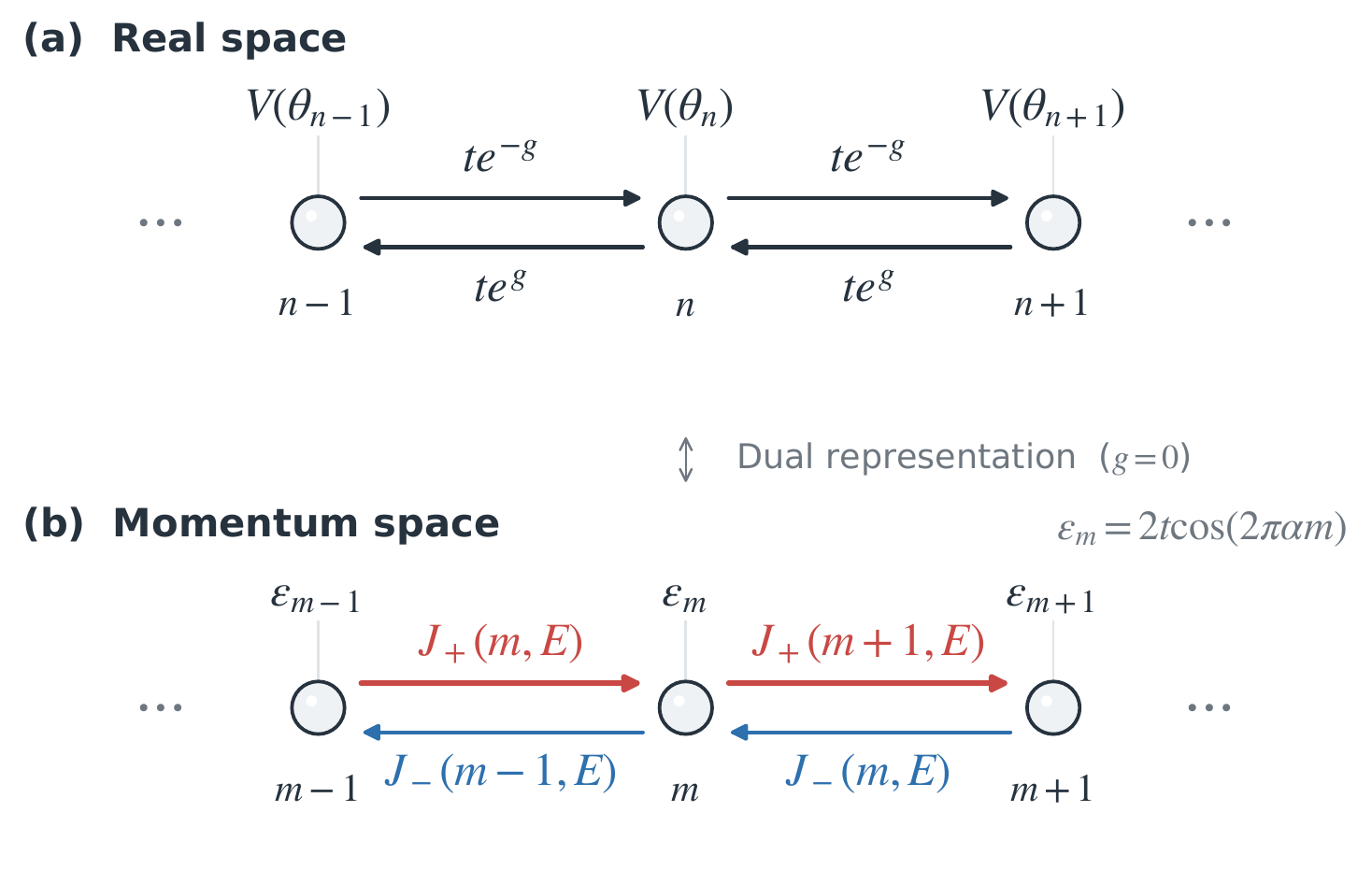}
    \caption{Potential and hopping schemes of the generalized non-Hermitian GPD model in real and momentum spaces. (a) Real-space chain with onsite potential $V(\theta_n)$ and hopping amplitudes $t e^{\pm g}$ in opposite directions. (b) Effective chain in momentum space at $g=0$, obtained after multiplying by the denominator of $V(\theta)$. $\varepsilon_m$ is the onsite term. $J_\pm(m,E)$ couple $u_{m\mp1}$ to site $m$ [see Eqs.~\eqref{eq:dualmodel} and \eqref{eq:dualhopping}]. The hopping profiles $J_\pm(m,E)$ depend on energy and could have distinct strengths.}
    \label{fig:model}
\end{figure}

Let $u_m$ be the wave amplitude at the $m$th lattice site in momentum space. Applying the dual transformation $\psi_n=\sum_m u_m e^{\ii m\theta_n}$ and multiplying both sides of Eq.~\eqref{eq:model} by the denominator of $V(z)$, we arrive at the Schr\"odinger equation in momentum space
\begin{equation}
J_+(m,E)u_{m-1}+J_-(m,E)u_{m+1}+\varepsilon_m u_m=Eu_m.
\label{eq:dualmodel}
\end{equation}
Here, $\varepsilon_m=2t\cos(2\pi\alpha m)$ is the potential at site $m$ in momentum space. The effective hopping coefficients $J_\pm(m,E)$ are
\begin{equation}
\begin{aligned}
J_+(m,E)&=p_+(E)-A\varepsilon_{m-1}=\lambda+\gamma+A(E-\varepsilon_{m-1}),\\
J_-(m,E)&=p_-(E)-B\varepsilon_{m+1}=\lambda-\gamma+B(E-\varepsilon_{m+1}).
\end{aligned}
\label{eq:dualhopping}
\end{equation}
Figure~\ref{fig:model}(b) illustrates the chain in momentum space. Because both hopping terms $J_\pm(m,E)$ depend on $E$ explicitly, Eq.~\eqref{eq:dualmodel} constitutes an energy-dependent dual problem of Eq.~\eqref{eq:model}.

In localized and extended phases related by the AAH/GPD duality~\cite{Aubry1980,Ganeshan2015,WangDuality2021,Goncalves2022}, extension/localization in momentum space accompanies localization/extension in real space. Changing the momentum-space hopping affects how strongly different momentum components are mixed and can therefore drive a real-space transition between extended and localized phases. In our generalized GPD model, the complex modulation makes the hopping $J_\pm(m,E)$ different in the two momentum-space directions. As the parameters or energy vary, the localization contribution associated with one direction can reach its transition threshold while the other remains below it. The two directional hopping processes can thus give rise to two distinct candidate mobility-edge contours in the complex-energy plane.

Moreover, both hopping terms in Eq.~\eqref{eq:dualhopping} contain the eigenenergy $E$, with different parameter combinations. Their relative strengths can therefore change across the complex-energy plane, allowing the dominant contribution to real-space localization to switch between the two directions. This is the physical origin of their competition. The resulting physical mobility edge can consequently be formed from selected portions of the two candidate contours. We next use the exact Lyapunov exponent to determine the two thresholds and their selection rule, and compare these predictions with numerical results.

% ============================================================
% II.3 EXACT RESULTS, FIGURE 2, AND FIGURE 3 VERIFICATION
% ============================================================
\subsection{Exact mobility-edge contours and numerical verification}
\label{sec:competition}

\begin{figure*}[!t]
\centering
\safeincludegraphics[width=0.94\textwidth]{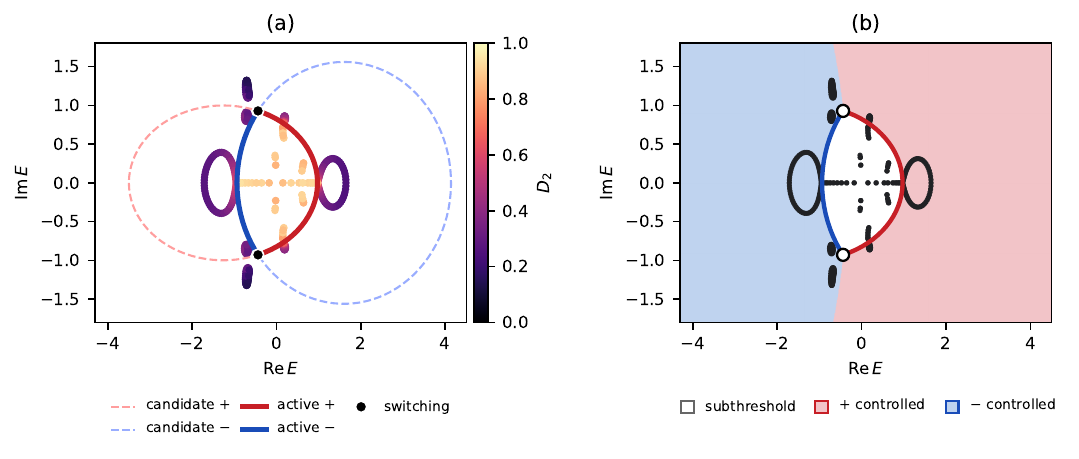}
\caption{Two localization channels and the selection of candidate boundaries. Parameters are $t=1$, $a=0.85$, $b=0.1$, $\lambda=0$, $\gamma=0.6$, $g=0$, $N=377$, $\alpha=233/377$, and $\varphi=0$. (a) Eigenenergies under PBC colored by the indicator $D_2$. Red and blue contours arise from the two analytical channel thresholds. Solid lines border the white region in (b) and satisfy Eq.~\eqref{eq:selection}. Dashed lines lie inside a colored region, where the stronger competing contribution sustains localization. (b) Analytical channel map: the white region has $\rho_+\le D$ and $\rho_-\le D$; red and blue regions indicate which channel dominates above threshold. Black dots mark the switching points $\rho_+=\rho_-=D$, where the selected red-white and blue-white boundary segments meet and exchange their controlling channel.}
\label{fig:competition}
\end{figure*}

For reciprocal hopping ($g=0$), the transfer matrix of Eq.~\eqref{eq:model} maps the wave components $(\psi_n,\psi_{n-1})^{\mathsf T}$ to $(\psi_{n+1},\psi_n)^{\mathsf T}$ and is given by
\begin{equation}
T_E(z)=\begin{pmatrix}[E-v(z)/d(z)]/t&-1\\1&0\end{pmatrix}.
\label{eq:transfer}
\end{equation}
Multiplying by $d(z)$ on both sides leads to the regularized transfer matrix $\widetilde T_E(z)=d(z)T_E(z)$. Using $Ed-v=E-p_+z-p_-z^{-1}$, we collect the terms proportional to $z$, $1$, and $z^{-1}$, which define the coefficient matrices $M_+$, $M_0$, and $M_-$ as
\begin{equation}
\begin{aligned}
\widetilde T_E(z)
&=z\underbrace{\begin{pmatrix}-p_+(E)/t&A\\-A&0\end{pmatrix}}_{M_+(E)}
+\underbrace{\begin{pmatrix}E/t&-1\\1&0\end{pmatrix}}_{M_0(E)}\\
&\quad+z^{-1}\underbrace{\begin{pmatrix}-p_-(E)/t&B\\-B&0\end{pmatrix}}_{M_-(E)}.
\end{aligned}
\label{eq:channel-decomposition}
\end{equation}
In the dual transformation, the coefficient of $e^{\ii m\theta_n}$ in $z\psi_n$ is $u_{m-1}$, whereas in $z^{-1}\psi_n$ it is $u_{m+1}$ with $z=e^{\ii\theta_n}$. These are the two shifts appearing in the effective hopping terms of Eq.~\eqref{eq:dualmodel}. We therefore refer to the $zM_+$ and $z^{-1}M_-$ contributions as the two \textit{Fourier channels}. Here, the indices $\tau=\pm$ label the positive and negative Fourier harmonics $z^{\pm1}$ of the regularized transfer matrix. Their dimensionless scales $\rho_\pm(E)$ are the spectral radii of $M_\pm(E)$, namely the largest eigenvalue moduli.

To quantify the two directional hopping processes, let $J_{\tau,\mathrm{geom}}(E)$ be the geometric mean of hopping magnitude $|J_\tau|$ over one period of its quasiperiodic modulation. Here, $\mathrm{geom}$ stands for geometric mean. $J_{\tau,\mathrm{geom}}$ is the nonnegative strength of the full hopping profile, whereas $J_\tau(m,E)$ is the coefficient at a particular dual site. Since localization is governed by exponential growth or decay under repeated wave-function transfer, we use a logarithmic phase average with $t$ the reference energy.

For $\tau=\pm$, we set $s_+=A$ and $s_-=B$. The exact relation between the hopping strengths and dimensionless channel scales is
\begin{equation}
\begin{aligned}
J_{\tau,\mathrm{geom}}(E)
&=t\exp\!\left[\int_0^{2\pi}\frac{\dd\theta}{2\pi}
\ln\left|\frac{p_\tau(E)-2ts_\tau\cos\theta}{t}\right|\right]=t\rho_\tau(E).
\end{aligned}
\label{eq:dualhoppingstrength}
\end{equation}
This identity follows from factoring the hopping profile using the channel eigenvalues and applying Jensen's formula. Explicit channel scales and phase-average calculations are given in Methods [Eqs.~\eqref{methods:characteristic}--\eqref{eq:hoppingradii}].

Multiplying the transfer matrix by $d(z)$ also changes its accumulated growth. At each site, the same scalar $d(e^{\ii\theta_n})$ multiplies the entire matrix, so a product of $N$ transfer matrices acquires the factor $\prod_{n=0}^{N-1}d(e^{\ii\theta_n})$. Let $\widetilde L(E,0)$ and $L_0(E)$ be the Lyapunov exponents of the regularized and original transfer matrices, respectively, evaluated with no imaginary phase shifts in $\theta$. Their difference is the mean logarithmic contribution of these scalar factors. Defining its exponential as $D$, we find
\begin{equation}
\begin{aligned}
\widetilde L(E,0)-L_0(E)
&=\lim_{N\to\infty}\frac1N\sum_{n=0}^{N-1}
\ln|d(e^{\ii\theta_n})|=\int_0^{2\pi}\frac{\dd\theta}{2\pi}
\ln|d(e^{\ii\theta})|\\
&\equiv\ln D.
\end{aligned}
\label{eq:denominator-normalization}
\end{equation}
In the pole-free regime, this gives
\begin{equation}
D=\frac{1+\sqrt{1-4AB}}{2}
=\frac{1+\sqrt{1-a^2+b^2}}{2}.
\label{eq:Ddef}
\end{equation}
Recovering the original Lyapunov exponent requires subtracting $\ln D$. Accordingly, the channel terms enter as $\ln(\rho_\pm/D)$, so their critical scale is $\rho_\pm=D$, or equivalently $J_{\pm,\mathrm{geom}}=tD$. Both channels share this normalization because they originate from the same denominator. The expression for $D$ is derived in Methods, Eq.~\eqref{methods:D}. Its extension to a complex phase shift is given in Supplementary Note~2.

For reciprocal hopping, Avila's global theory gives the result of Lyapunov exponent for the original transfer matrix as
\begin{equation}
L_0(E)=\max\!\left\{0,\ln\frac{\rho_+(E)}{D},
\ln\frac{\rho_-(E)}{D}\right\}.
\label{eq:exactLE}
\end{equation}
The derivation is given in Methods. If both $\rho_\pm\leq D$, the $L_0(E)$ vanishes. If either one of $\rho_\pm$ exceeds $D$, the $L_0(E)$ is positive. When both $\rho_\pm>D$, the larger one determines the Lyapunov exponent and controls the localization transition.

The individual thresholds $\rho_+=D$ and $\rho_-=D$, or $J_{+,\mathrm{geom}}=tD$ and $J_{-,\mathrm{geom}}=tD$, give the candidate curves for $a\pm b\ne0$ (see Supplementary Note~2 for derivations), i.e.,
\begin{equation}
E_\pm^{(0)}(\phi)=-\frac{2(\lambda\pm\gamma)}{a\pm b}-\frac{2t}{a\pm b}\left[De^{\ii\phi}
+\frac{(a\pm b)^2}{4D}e^{-\ii\phi}\right].
\label{eq:contours}
\end{equation}
The angle $\phi\in[0,2\pi)$ traces each complex-energy curve and differs from the lattice phase $\varphi$. These curves are ellipses for real parameters, with centers and shapes determined by the different parameter combinations in $p_\pm(E)$. If $a+b=0$ or $a-b=0$, the corresponding radius is independent of energy and must be treated directly rather than with this ellipse formula.

Figure~\ref{fig:competition} compares these analytical boundaries with numerical spectra and illustrates how the two contributions select the physical mobility edge. In panel (a), the red and blue curves represent the thresholds $\rho_+=D$ and $\rho_-=D$. The eigenstates under PBC are colored by the indicator $D_2$: values closer to zero indicate localization, while values closer to one indicate extended states. Panel (b) shows which contribution dominates above the threshold and explains why only certain portions of the two curves form the transition boundary.

To understand which parts of these curves can describe a localization transition, we start from the white region in Fig.~\ref{fig:competition}(b). There, both contributions are at or below the threshold, and all eigenstates have $L_0=0$. In the red region, the $+$ branch exceeds the threshold and becomes dominant, while the $-$ branch does so in the blue region. Therefore, the boundary between the white and a colored region is where the spectral Lyapunov exponent changes between zero and a positive value.

The solid red segment in Fig.~\ref{fig:competition}(b) divides the white and red regions. On this segment, the $+$ branch just reaches the threshold, $\rho_+=D$, while the $-$ one is at or below it. Entering the red region makes the $+$ branch sustain localization. Similarly, the solid blue segment separates the white and blue regions: the $-$ branch reaches the threshold while the $+$ one does not exceed it. These two cases give the selection rule
\begin{equation}
\rho_\tau(E)=D,\qquad \rho_{-\tau}(E)\le D,
\qquad\tau=\pm.
\label{eq:selection}
\end{equation}
The spectrum in panel (a) shows where these selected boundaries are reached by eigenstates, and the changes in $D_2$ track the associated change in spatial localization.

The dashed segments in Fig.~\ref{fig:competition}(a) have a different meaning. Consider the red dashed curve inside the blue region. The $+$ branch reaches its own threshold there, but the stronger $-$ branch has already exceeded the threshold. Eigenstates therefore have $L_0(E)=\ln[\rho_-(E)/D]>0$ on the red dashed curve and remain localized on both sides of it. Crossing this curve changes whether the weaker contribution is above or below its own threshold, but does not change the localized character of the state. This is why that part of the red ellipse is suppressed as a mobility edge. A blue dashed segment inside the red region has the same interpretation with the labels exchanged. Fig.~\ref{fig:competition}(b) therefore explains why the spectral crossing of a dashed curve in Fig.~\ref{fig:competition}(a) needs not produce a localization transition.

Finally, we follow the edge of the white region through a black dot in Fig.~\ref{fig:competition}(b). The boundary changes from a solid red to a solid blue segment, or vice versa. At this junction, denoted by $E_s$, the contributions of both branches are exactly at the threshold:
\begin{equation}
\rho_+(E_s)=\rho_-(E_s)=D.
\label{eq:switchgeneral}
\end{equation}
On one side of the dot, the $+$ branch determines the transition boundary, while the $-$ one does on the other side. This is the exchange of control between the two candidate ellipses. The analytical map provides this geometric picture, while the eigenstates in Fig.~\ref{fig:competition}(a) reveal its spectral pattern.

\begin{figure}[!t]
\centering
\safeincludegraphics[width=0.88\columnwidth]{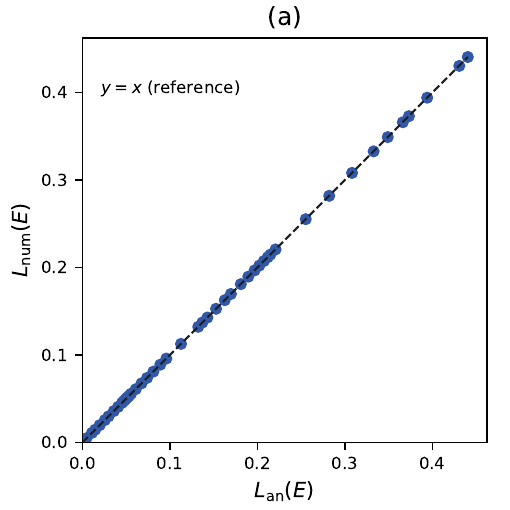}
\caption{Numerical test of the exact Lyapunov exponent. Parameters are $t=1$, $a=0.85$, $b=0.1$, $\lambda=0$, $\gamma=0.6$, $g=0$, and $\varphi=0$. The $60$ test energies are sampled from the periodic chain with $N=377$ and $\alpha=233/377$. At each energy, $L_{\mathrm{num}}$ is obtained from $30000$ normalized transfer-matrix steps at $\alpha=(\sqrt{5}-1)/2$, while $L_{\mathrm{an}}$ is evaluated from Eq.~\eqref{eq:exactLE}. The dashed line indicates $L_{\mathrm{num}}=L_{\mathrm{an}}$.}
\label{fig:LEvalidation}
\end{figure}

Figure~\ref{fig:LEvalidation} verifies the analytical results. It compares Eq.~\eqref{eq:exactLE} with Lyapunov exponents obtained by multiplying the original transfer matrices at sampled eigenenergies of finite Fibonacci approximants, yielding perfect agreements.

% ============================================================
% II.4 SPECIAL MODELS AND CLASSIFICATION OF DUAL HOPPING
% ============================================================
\subsection{Dual-hopping-based classification of AAH and GPD models}\label{sec:limits}
\begin{figure*}[!t]
\centering
% Full-width comparison schematic.
% Plot geometric-mean hopping strengths J_{+,geom} and J_{-,geom} versus energy.
\safeincludegraphics[width=0.90\textwidth]{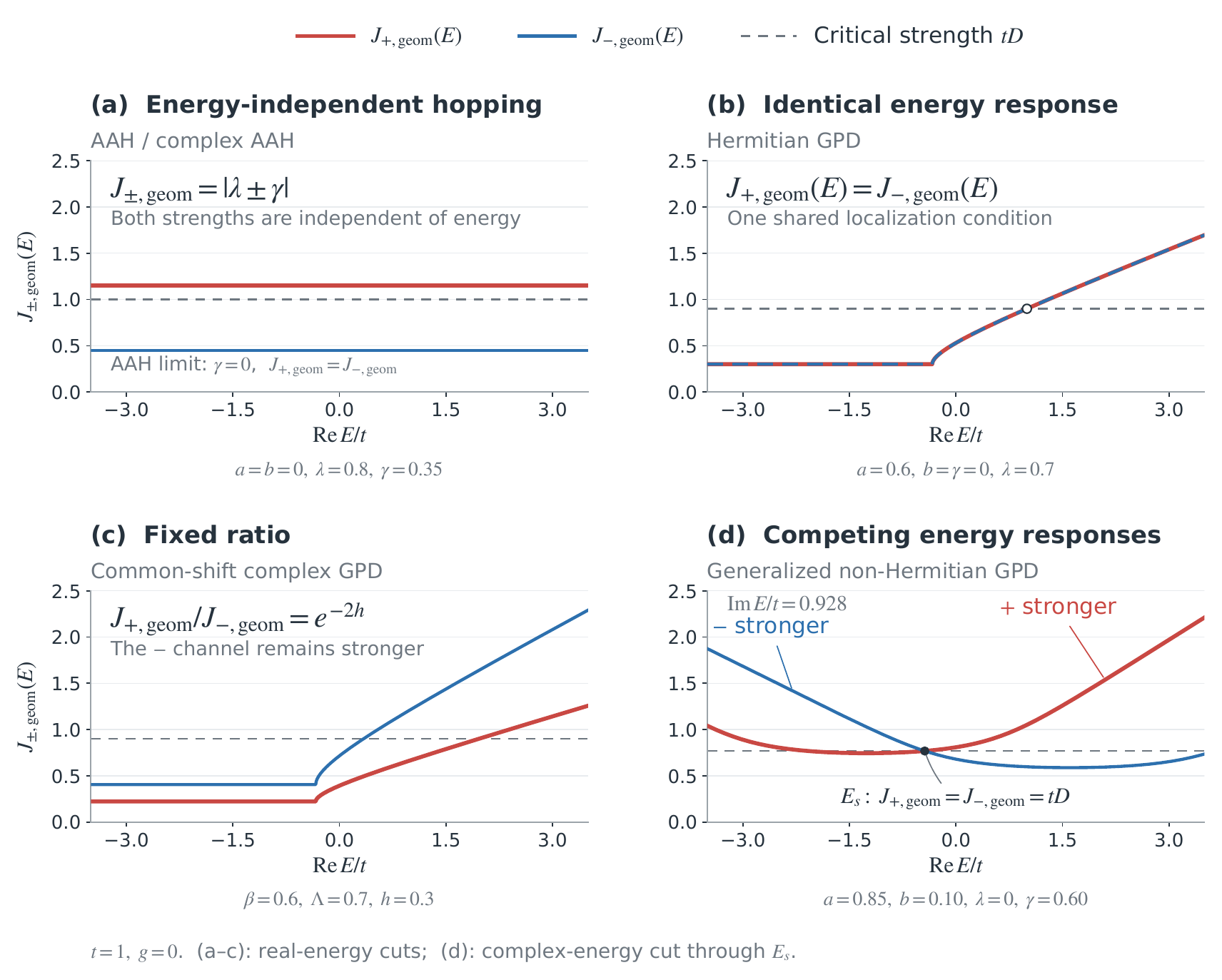}
\caption{Classification of directional momentum-space hopping. The curves schematically represent the geometric-mean hopping strengths $J_{+,\mathrm{geom}}(E)$ and $J_{-,\mathrm{geom}}(E)$ of the reciprocal reference problem. The horizontal threshold is $tD$. For complex energies, the horizontal axis follows a specified path in the complex plane. (a) Energy-independent strengths: equal in the AAH model and generally unequal in the complex AAH-type chains. (b) Identical energy-dependent strengths: the Hermitian GPD model and the related construction of Ref.~\cite{Wang2023}. The reciprocal reference of the nonreciprocal GPD model also belongs to this class. (c) Energy-dependent strengths with a fixed ratio $J_{+,\mathrm{geom}}/J_{-,\mathrm{geom}}=e^{-2h}$ in the common-shift complex GPD model. (d) Independently tunable channel strengths in our generalized model, allowing the order of $J_{\pm,\mathrm{geom}}$ to reverse. The physical boundary is selected by Eq.~\eqref{eq:selection}. An exchange of control along that boundary occurs when both strengths reach $tD$, as specified by Eq.~\eqref{eq:switchgeneral}.}
\label{fig:modelcomparison}
\end{figure*}

The exact results above allow us to classify various known AAH and GPD models within the same framework according to their directional hopping strengths $J_{\pm,\mathrm{geom}}(E)$, as illustrated in Figure~\ref{fig:modelcomparison}.  We organize these models into three general groups, i.e., (i) both $J_{\pm,\mathrm{geom}}$ are $E$-independent, (ii) $J_{+,\mathrm{geom}}(E)=J_{-,\mathrm{geom}}(E)$ for all $E$, and (iii) the ratio $J_{+,\mathrm{geom}}(E)/J_{-,\mathrm{geom}}(E)$ is fixed. We apply Eqs.~\eqref{eq:exactLE}, \eqref{eq:selection} and use $\rho_\tau=J_{\tau,\mathrm{geom}}/t$ to explain the localization behavior in each case.

\textit{Group (i).}
In the AAH limit $a=b=\gamma=g=0$, both dual hoppings are equal to $\lambda$, so $J_{+,\mathrm{geom}}=J_{-,\mathrm{geom}}=|\lambda|$. With $D=1$, Eq.~\eqref{eq:exactLE} predicts the common transition at $|\lambda|=t$. The complex AAH-type chain with $a=b=g=0$ and $\gamma\ne0$ instead has unequal hoppings $J_\pm=\lambda\pm\gamma$ and $J_{\pm,\mathrm{geom}}=|\lambda\pm\gamma|$, but neither depends on $E$. Our theory predicts the transition at $\max\{|\lambda+\gamma|,|\lambda-\gamma|\}=t$. The horizontal curves in Fig.~\ref{fig:modelcomparison}(a) illustrate this energy independence: all eigenstates share the same localization threshold, showing no mobility edges.

\textit{Group (ii).}
In the GPD limit $b=\gamma=g=0$, both hopping profiles have the form $\lambda+(a/2)(E-\varepsilon)$ with identical geometric means $J_{+,\mathrm{geom}}(E)=J_{-,\mathrm{geom}}(E)$, as represented by the overlapping curves in Fig.~\ref{fig:modelcomparison}(b). The common threshold $J_{\pm,\mathrm{geom}}=tD$ gives the unique localization condition $|aE+2\lambda|=2t$. Its two algebraic branches come from the same localization scale. A branch describes a physical mobility edge only where it separates spectral states with different localization character. Adding real-space nonreciprocity while keeping $b=\gamma=0$ preserves the identical strengths of the reciprocal reference problem and shifts their common threshold to $tDe^{|g|}$~\cite{Liu2021Dual}.
The generalized Hermitian GPD model of Ref.~\cite{Wang2023} adds a constant numerator term. This changes the spectrum without separating the two directional energy responses, placing it in the same class as Fig.~\ref{fig:modelcomparison}(b). In the bounded-potential regime, the two branches reported there are algebraic solutions of one localization condition, and the spectrum selects the physical branch. They therefore arise from a common localization scale rather than the two competing directional contributions considered here.

\textit{Group (iii).}
A common non-Hermitian GPD construction applies the same imaginary phase shift $\theta\to\theta+\ii h$ to the numerator and denominator of $V(\theta)$ with $h\in{\mathbb R}$ \cite{Liu2021c}. The $\Lambda$ and $\beta$ in Fig.~\ref{fig:modelcomparison}(c) specify the numerator amplitude and denominator modulation in $V_h(\theta)=2\Lambda\cos(\theta+\ii h)/[1-\beta\cos(\theta+\ii h)]$, where $\Lambda$ is related to the $\lambda$ in Eq.~\eqref{eq:potential} by $\lambda=\Lambda\cosh h$. The two dual hopping profiles then share the same energy-dependent form and multiplied by fixed factors $e^{-h}$ and $e^{h}$ at their respective shifted sites, yielding $J_{+,\mathrm{geom}}(E)=e^{-2h}J_{-,\mathrm{geom}}(E)$ as illustrated in Fig.~\ref{fig:modelcomparison}(c). Changing $E$ maintains their ratio fixed. For $h>0$ ($h<0$) the $-$ ($+$) direction remains stronger, while for $h=0$ they coincide. Applying Eq.~\eqref{eq:selection}, only the stronger direction can supply the localization boundary. Two possible threshold contours can thus exist without exchanging control.

As a key advantage, our model allows $a$, $b$, $\lambda$, and $\gamma$ to vary independently within the pole-free regime. The two hoppings in Eq.~\eqref{eq:dualhopping} thus contain independently adjustable coefficients of $E$ and offsets, rather than sharing an identical profile or a fixed proportionality. Their geometric-mean strengths can exchange dominance as energy varies, as illustrated in Fig.~\ref{fig:modelcomparison}(d). The larger one determines the dominant localization contribution in Eq.~\eqref{eq:exactLE}. Each direction can offer the selected boundary where $J_{\tau,\mathrm{geom}}=tD$ while $J_{-\tau,\mathrm{geom}}\le tD$. When they both reach $tD$, the selected segments can join and exchange control, as illustrated in Fig.~\ref{fig:competition}. Solving the general model reveals this competition, which is absent in the special groups (i)--(iii) above. Table~\ref{tab:comparison} summarizes our classification, with additional parameter reductions given in Supplementary Note~1.

\begin{table*}[!t]
\caption{AAH and GPD models classified by the energy dependence of their dual hopping, corresponding to Fig.~\ref{fig:modelcomparison}. The listed models represent limiting cases or constrained subfamilies of Eqs.~\eqref{eq:model} and \eqref{eq:potential}, except for the related model of Ref.~\cite{Wang2023}. The geometric-mean hopping strengths $J_{\pm,\mathrm{geom}}$ are defined in Eq.~\eqref{eq:dualhoppingstrength} for the reciprocal reference problem. Their thresholds are $tD$ at $g=0$ and $tDe^{|g|}$ for the nonreciprocal criterion. Counts of physical mobility edges refer to regimes where localized and extended states coexist in the spectrum. Two algebraic branches of one localization condition are distinguished from the two separate directional candidate curves of the present model.}
\label{tab:comparison}
\centering
\fontsize{7.7}{8.8}\selectfont
\renewcommand{\arraystretch}{1.16}
\setlength{\tabcolsep}{4pt}
\begin{tabularx}{\textwidth}{@{}p{2.5cm}p{2.95cm}p{3.15cm}Y@{}}
\toprule
Model & Parameter choice or potential & Relation between hopping strengths & Mobility-edge behavior \\
\midrule
\multicolumn{4}{@{}l}{\textit{Group (i) [Fig.~\ref{fig:modelcomparison}(a)]}}\\
AAH & $a=b=\gamma=g=0$ & $J_{+,\mathrm{geom}}=J_{-,\mathrm{geom}}=|\lambda|$ & No mobility edge; all energies share the same transition threshold $|\lambda|=t$ \\
Complex AAH-type chain & $a=b=g=0$, $\gamma\ne0$ & $J_{\pm,\mathrm{geom}}=|\lambda\pm\gamma|$ & No mobility edge; unequal but energy-independent contributions give the same localization threshold at all energies \\
\addlinespace
\multicolumn{4}{@{}l}{\textit{Group (ii) [Fig.~\ref{fig:modelcomparison}(b)]}}\\
Hermitian GPD~\cite{Ganeshan2015} & $b=\gamma=g=0$ & $J_{+,\mathrm{geom}}=J_{-,\mathrm{geom}}$ & One physical mobility edge; identical directional contributions give one localization condition, and the spectrum selects its physical root \\
Nonreciprocal GPD & $b=\gamma=0$, $g\ne0$ & Equal $J_{\pm,\mathrm{geom}}$ in the reciprocal reference & One mobility-edge boundary; hopping asymmetry shifts the common threshold of the identical reciprocal contributions \\
Generalized GPD~\cite{Wang2023} & Additional constant numerator term & Equal $J_{\pm,\mathrm{geom}}$; one common condition & One physical mobility edge in the bounded-potential regime; the two reported algebraic branches solve the same localization condition, and the spectrum selects the physical branch \\
\addlinespace
\multicolumn{4}{@{}l}{\textit{Group (iii) [Fig.~\ref{fig:modelcomparison}(c)]}}\\
Common-shift complex GPD~\cite{Liu2021c} & Shared shift $\theta\to\theta+\ii h$ & $J_{+,\mathrm{geom}}=e^{-2h}J_{-,\mathrm{geom}}$ & One mobility-edge boundary; a fixed ratio keeps the same channel stronger at every energy \\
\addlinespace
\multicolumn{4}{@{}l}{\textit{General case [Fig.~\ref{fig:modelcomparison}(d)]}}\\
Our model & Independent $a,b,\lambda,\gamma,g$ & $J_{\pm,\mathrm{geom}}(E)$ can exchange order & Two directional candidate curves; selected portions reached by the spectrum form the physical mobility edge, with the controlling channel able to switch between them \\
\bottomrule
\end{tabularx}
\end{table*}

% ============================================================
% II.5 NONRECIPROCITY, FIGURE 5, AND THE FOUR STAR MARKERS
% ============================================================
\subsection{Nonreciprocal hopping and shifted mobility edges}\label{sec:nonreciprocity}
Finally, we switch on the hopping asymmetry in real-space by setting $g\ne0$. In the dual equation, $\varepsilon_m$ becomes $2t\cos(2\pi\alpha m-\ii g)$, while the form of Eq.~\eqref{eq:dualhopping} is preserved. Physically, asymmetric hopping favors motion in one direction and competes with the confinement of quasiperiodic potential. Greater hopping asymmetry demands stronger disorder to localize a state spatially.

The imaginary gauge transformation for an open chain offers a way to understand this competition. It shows how asymmetric hopping changes the spatial profile of the wave function, motivating a comparison between $|g|$ and the $L_0(E)$ of the reciprocal reference model. For the system under PBC, localization persists when $L_0(E)>|g|$. The threshold $L_0(E)=|g|$ marks the balance between confinement and hopping asymmetry. This is the established imaginary gauge delocalization criterion of Hatano and Nelson \cite{Hatano1996,Hatano1998}. Applying Eq.~\eqref{eq:exactLE} to the reciprocal localized spectrum then yields
\begin{equation}
\max\!\left\{\rho_+(E),\rho_-(E)\right\}=R_g,
\qquad R_g\equiv De^{|g|}.
\label{eq:Rg}
\end{equation}
The comparison is thus unchanged in form, and we only need to replace the $D$ by $R_g$ in the candidate ellipses and selection rule. Each channel controls a selected segment only if it reaches $R_g$ while the other does not. Increasing $|g|$ raises the disorder strength required to counteract hopping asymmetry, shifts candidate boundaries, and changes their intersections. The scales $\rho_\pm$ remain those of the reciprocal reference problem.

\begin{figure}[!t]
\centering
\safeincludegraphics[width=0.92\linewidth]{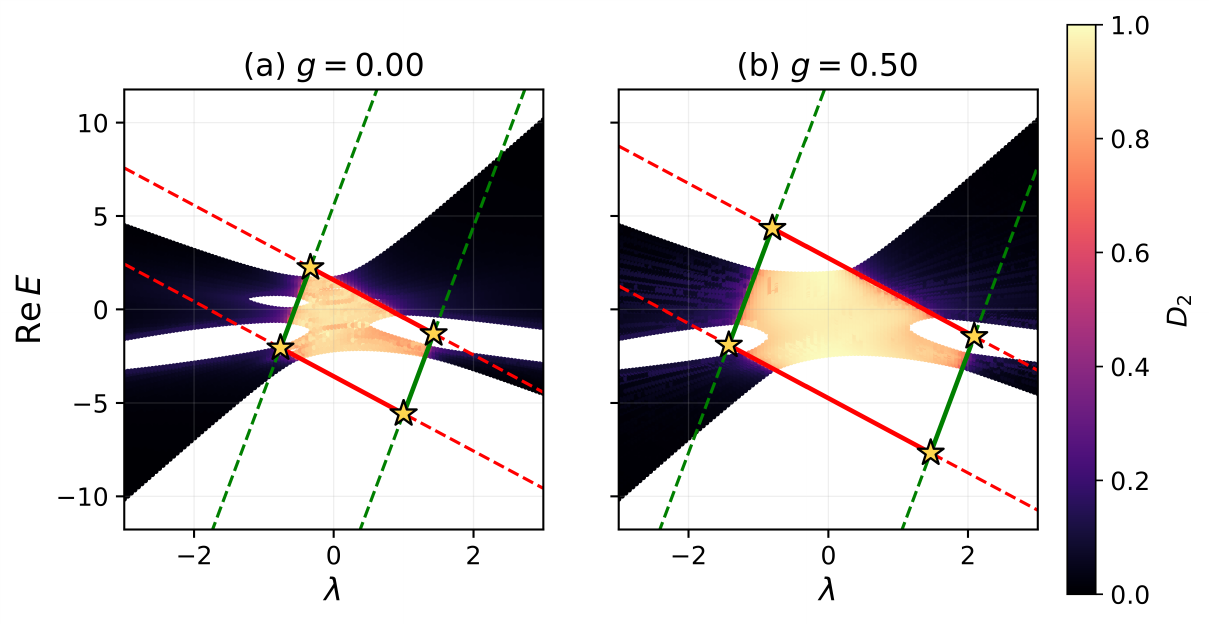}
\caption{Nonreciprocity shifts the channel thresholds. Real parts of spectra under PBC for (a) $g=0$ and (b) $g=0.5$, with $t=1$, $N=377$, $\alpha=233/377$, $\gamma=0.5$, $a=0.4$, and $b=0.6$. Colors encode the fractal dimension $D_2$. Red and green curves show how the real-axis intercepts of the candidate boundaries for the $+$ and $-$ channels, respectively, vary with $\lambda$. Solid (dashed) parts satisfy (fail) the selection inequality. Filled stars mark the four possible algebraic coincidences of real-axis intercepts of the two candidates, computed with $R=D$ in (a) and $R=R_g$ in (b) [see Supplementary Eqs.~(S31) and (S32) for details.]}
\label{fig:phase}
\end{figure}

Figure~\ref{fig:phase} compares the real part of PBC spectra for $g=0$ and $g=0.5$ as $\lambda$ varies. The colors encode $D_2$. Solid curves pass the channel comparison and dashed curves are suppressed by the other channel. The four stars mark the coincidences of the two candidates' real-axis intercepts. Their coordinates are obtained by equating these intercepts, using $D$ in Fig.~\ref{fig:phase}(a) and $R_g$ in Fig.~\ref{fig:phase}(b), as detailed in Supplementary Eqs.~(S31) and (S32). At each star, both channels simultaneously reach the threshold, providing analytical reference positions for comparing the candidate-branch structure with the spectrum.

\section{Discussion}

Our results reveal how complex quasiperiodic modulations can produce mobility edges controlled by different localization contributions. In momentum space, these contributions are associated with hopping along opposite directions. The generalized GPD potential makes both hopping terms energy dependent, but in different ways, so the direction that controls localization can change across the spectrum. The mobility edge can then combine portions of two different critical curves. Our model also incorporates representative AAH and GPD models as specific limits in which the two effective hopping terms are identical, differ only by a fixed factor, or are independent of energy. These extra constraints prevent the switch of control found in our general model.

The found competition implies that reaching the threshold of one Fourier channel does not necessarily imply a localization transition, as the other channel may dominate localization. The physical mobility edge is therefore formed by the selected spectral portions of the two candidate curves, as illustrated by the solid segments in Fig.~\ref{fig:competition}.

Beyond clarifying how multiple analytical contours jointly determine localization, our theoretical framework opens several avenues for future explorations. Potential experimental platforms include photonic lattices with engineered gain and loss, acoustic resonator arrays, ultracold atoms in bichromatic optical lattices, and active electrical circuits. Photonic quantum walks and electrical circuits have already demonstrated key ingredients of non-Hermitian lattice physics~\cite{Weidemann2022,Helbig2020}. Implementing the quasiperiodic modulation of the present model in such platforms could allow the predicted channel switching to be probed through spatially resolved intensity, transmission, or electrical-response measurements. Theoretically, extending our analysis to open boundaries would reveal how competing momentum-space channels interact with the non-Hermitian skin effect. Generalizing the mechanism to long-range hopping and interacting systems provides further directions for future work.

\section{Methods}
\label{sec:methods}

\setcounter{equation}{0}
\renewcommand{\theequation}{M\arabic{equation}}
\renewcommand{\theHequation}{M.\arabic{equation}}

\subsection{Analytical representation and channel scales}
\label{methods:channels}
For reciprocal hopping ($g=0$), we write $d(\theta)=d(e^{\ii\theta})$. The transfer matrix and its regularized form are given by
\begin{equation}
\begin{alignedat}{2}
&T_E(\theta) &&{}=\begin{pmatrix}[E-V(\theta)]/t&-1\\1&0\end{pmatrix},\\
&\widetilde T_E(\theta) &&{}=d(\theta)T_E(\theta),\qquad\det T_E=1.
\end{alignedat}
\label{methods:transfer}
\end{equation}
We use $\epsilon\in\mathbb R$ for an imaginary phase shift. In our parameter regime, the denominator of $V(\theta)$ is nonzero for all $\theta\in\mathbb R$ but may vanish after the shift $\theta\to\theta+\ii\epsilon$. The regularized transfer matrix has no such poles, and its scalar normalization can be removed exactly. Let $T^{(N)}_{E,\epsilon}$ and $\widetilde T^{(N)}_{E,\epsilon}$ be the ordered products of $N$ transfer matrices along the chain. Since the scalar factors commute with all matrices, the products satisfy
\begin{equation}
\widetilde T^{(N)}_{E,\epsilon}(\theta)
=\left[\prod_{j=0}^{N-1}d(\theta+2\pi j\alpha+\ii\epsilon)\right]T^{(N)}_{E,\epsilon}(\theta).
\label{methods:product-normalization}
\end{equation}
On a phase circle without zeros in denominators, taking logarithmic norms and averaging over $\theta$ gives the same scalar average for each factor. After division by $N$ and taking the limit $N\to\infty$, the phase-averaged Lyapunov exponents obey
\begingroup
\allowdisplaybreaks[4]
\begin{alignat}{2}
&L_0(E,\epsilon) &&{}=\lim_{N\to\infty}\frac1N\int_0^{2\pi}\frac{\dd\theta}{2\pi}
\ln\|T^{(N)}_{E,\epsilon}(\theta)\|,\notag\\[3pt]
&\widetilde L(E,\epsilon) &&{}=\lim_{N\to\infty}\frac1N\int_0^{2\pi}\frac{\dd\theta}{2\pi}
\ln\|\widetilde T^{(N)}_{E,\epsilon}(\theta)\|,\notag\\[3pt]
&L_0(E,\epsilon) &&{}=\widetilde L(E,\epsilon)-I_d(\epsilon),\notag\\*[3pt]
&I_d(\epsilon) &&{}=\int_0^{2\pi}\frac{\dd\theta}{2\pi}\ln|d(\theta+\ii\epsilon)|,\notag\\*[3pt]
&\ln D &&{}=I_d(0).
\label{methods:normalization}
\end{alignat}
\endgroup
Each regularized transfer matrix differs by a scalar $d$, so the result picks up the product of these scalar factors. Its logarithmic average $I_d(\epsilon)$ gives the subtraction in Eq.~\eqref{methods:normalization}. For $A\ne0$, let $z_1$ and $z_2$ be the roots of $Az^2-z+B=0$, so that $d(z)=-A(z-z_1)(z-z_2)/z$. Because $|A|+|B|=\max\{|a|,|b|\}<1$, the term $-z$ dominates $Az^2+B$ on the unit circle. Therefore, exactly one root is inside and the other is outside. Jensen's formula then gives
\begin{equation}
\begin{alignedat}{2}
&z_{1,2} &&{}=\frac{1\pm\sqrt{1-4AB}}{2A},\\
&D &&{}=|A|\prod_{j=1}^{2}\max\{1,|z_j|\}=\frac{1+\sqrt{1-4AB}}{2}.
\end{alignedat}
\label{methods:D}
\end{equation}
For $A=0$, $|B|<1$ gives $D=1$, consistent with the same expression. The locations of two roots also imply $D>|A|,|B|$, as required by the contour construction in Results. Since no roots cross the unit circle for a sufficiently small phase shift, $I_d(\epsilon)=\ln D$ near $\epsilon=0$. Supplementary Note~2 gives $I_d(\epsilon)$ for arbitrary phase shifts.

Substituting $Ed-v=E-p_+z-p_-z^{-1}$ and $d=1-Az-Bz^{-1}$ into the regularized transfer matrix gives the Fourier decomposition
\begin{equation}
\begin{alignedat}{2}
&\widetilde T_E(z)
&&{}=\begin{pmatrix}(Ed-v)/t&-d\\d&0\end{pmatrix}=z\begin{pmatrix}-p_+(E)/t&A\\-A&0\end{pmatrix}
\\
& &&{}+\begin{pmatrix}E/t&-1\\1&0\end{pmatrix}+z^{-1}\begin{pmatrix}-p_-(E)/t&B\\-B&0\end{pmatrix}\\
& &&{}=zM_+(E)+M_0(E)+z^{-1}M_-(E).
\end{alignedat}
\label{methods:channel-decomposition}
\end{equation}
The upper-left entries of $M_\pm$ are the coefficients $-p_\pm(E)/t$ from $Ed-v$. Their off-diagonal entries $A,-A$ and $B,-B$ come from the multiplication by $d(z)$. Thus, $M_\pm$ are the coefficient matrices of the two Fourier harmonics. The complete regularized matrix has the determinant $d^2(z)$.

With $s_+=A$ and $s_-=B$ as defined in Results, let $\eta_{\tau,\mu}$ ($\mu=\pm1$) be the two eigenvalues of $M_\tau$ for $\tau=\pm$. They are the roots of equation
\begin{equation}
\det(\eta I-M_\tau)
=\eta^2+\frac{p_\tau(E)}{t}\eta+s_\tau^2=0.
\label{methods:characteristic}
\end{equation}
Consequently, their larger modulus gives the channel scale:
\begin{equation}
\begin{alignedat}{2}
&M_\tau(E) &&{}=\begin{pmatrix}-p_\tau(E)/t&s_\tau\\-s_\tau&0\end{pmatrix},\\
&\eta_{\tau,\mu}(E) &&{}=\frac{-p_\tau(E)+\mu\sqrt{p_\tau^2(E)-4t^2s_\tau^2}}{2t},\\
&\rho_\tau(E) &&{}=\max_{\mu=\pm1}|\eta_{\tau,\mu}(E)|,\\
&\eta_{\tau,+1}+\eta_{\tau,-1} &&{}=-p_\tau(E)/t,\qquad \eta_{\tau,+1}\eta_{\tau,-1}=s_\tau^2.
\end{alignedat}
\label{methods:radii}
\end{equation}
Changing the square-root branch only interchanges the two eigenvalues and does not affect $\rho_\tau$. Repeated roots are also included in this definition.

For $g=0$, we denote the logarithmic average of the magnitude of hopping in Eq.~\eqref{eq:dualhopping} over one period of its quasiperiodic modulation by $\mathcal{I}_\tau(E)$, i.e.,
\begin{equation}
\mathcal I_\tau(E)=\int_0^{2\pi}\frac{\dd\theta}{2\pi}
\ln\left|\frac{p_\tau(E)-2ts_\tau\cos\theta}{t}\right|.
\label{methods:hopping-mean}
\end{equation}
For $s_\tau\ne0$, we set $z=e^{\ii\theta}$. The sum and product of eigenvalues in Eq.~\eqref{methods:radii} give the factorization
\begin{equation}
\frac{p_\tau-ts_\tau(z+z^{-1})}{t}
=-s_\tau z^{-1}
\prod_{\mu=\pm1}\left(z+\frac{\eta_{\tau,\mu}}{s_\tau}\right).
\label{methods:hopping-factorization}
\end{equation}
Here, $z^{-1}$ has unit modulus on the unit circle. Jensen's formula gives the phase average of the logarithm of each remaining factor as $\ln\max\{1,|\eta_{\tau,\mu}|/|s_\tau|\}$. Hence
\begin{equation}
\mathcal I_\tau(E)
=\ln|s_\tau|+
\sum_{\mu=\pm1}\ln\max\left\{1,\frac{|\eta_{\tau,\mu}|}{|s_\tau|}\right\}=\ln\rho_\tau(E).
\label{methods:hopping-Jensen}
\end{equation}
To see the last equality, the product of the eigenvalue moduli is $s_\tau^2$. The larger modulus is therefore at least $|s_\tau|$, and the smaller is at most $|s_\tau|$. Only the larger eigenvalue contributes a possibly nonzero Jensen term, namely $\ln(\rho_\tau/|s_\tau|)$. This argument also covers equal eigenvalue moduli. Isolated zeros of the hopping profile give integrable logarithmic singularities.

Exponentiating and restoring the hopping unit $t$ yields the geometric-mean strength defined in Eq.~\eqref{eq:dualhoppingstrength}, i.e.,
\begin{equation}
\begin{alignedat}{2}
&J_{\tau,\mathrm{geom}}(E) &&{}=t\rho_\tau(E)=t\exp[\mathcal I_\tau(E)]\\[3pt]
& &&{}=t\exp\!\left[\int_0^{2\pi}\frac{\dd\theta}{2\pi}
\ln\left|\frac{p_\tau(E)-2ts_\tau\cos\theta}{t}\right|\right]\\[3pt]
& &&{}=\frac12\max_{\mu=\pm1}\left|-p_\tau(E)
+\mu\sqrt{p_\tau^2(E)-4t^2s_\tau^2}\right|.
\end{alignedat}
\label{eq:hoppingradii}
\end{equation}
The last line is $t$ times the spectral radius in Eq.~\eqref{methods:radii}. Thus, $J_{\pm,\mathrm{geom}}(E)=t\rho_\pm(E)$ are the geometric-mean magnitudes of the full directional effective hopping profiles. For $s_\tau=0$, the hopping profile is constant, $\mathcal I_\tau=\ln(|p_\tau|/t)$, and the hopping strength reduces to $|p_\tau|=t\rho_\tau$. If $p_\tau=0$ as well, $\mathcal I_\tau=-\infty$ and the strength is zero, so the same identity includes a vanishing channel.

\subsection{Complex phase shifts and channel growth scales}\label{methods:proof}
To separate the contributions of two Fourier channels, we first shift $\theta\to\theta+\ii\epsilon$, so that $|z|=e^{-\epsilon}$. For large negative (positive) $\epsilon$, the term $zM_+$ ($z^{-1}M_-$) dominates the Laurent expansion. Removing the corresponding scalar power of $z$ leaves, in each limit, the constant matrix $M_+$ or $M_-$. The powers removed account for the linear $-\epsilon$ or $+\epsilon$ term in the Lyapunov exponent. Since a constant matrix grows at the logarithm of its spectral radius, the two limits are
\begin{equation}
\begin{alignedat}{2}
&\widetilde L(E,\epsilon) &&{}=-\epsilon+\ln\rho_+(E)+o(1),\quad \epsilon\to-\infty,\\[3pt]
&\widetilde L(E,\epsilon) &&{}=+\epsilon+\ln\rho_-(E)+o(1),\quad \epsilon\to+\infty.
\end{alignedat}
\label{methods:asymptotics}
\end{equation}
These asymptotic expressions apply to nonzero channel radii. The $o(1)$ terms vanish in the stated limits. A vanishing channel does not contribute finite logarithmic bounds, as detailed in the Supplementary Note~3. The convexity of the regularized Lyapunov exponent makes the two limiting straight lines lower bounded at $\epsilon=0$~\cite{Jitomirskaya2012,LiAsym2024,Tong2025}. The determinant-one condition also implies $L_0(E)\ge0$. Together, these facts give
\begin{equation}
\widetilde L(E,0)\ge\max\{\ln D,\ln\rho_+(E),\ln\rho_-(E)\}.
\label{methods:bound}
\end{equation}

To show that the bound is attained for spectral energies, we apply the quantization of acceleration in Avila's global theory~\cite{Avila2015}. The denominator remains nonzero for $\theta+\ii\epsilon$ in a strip $|\epsilon|<\delta$ around $\epsilon=0$, for some $\delta>0$, so the original matrix defines an analytic SL$(2,\mathbb{C})$ cocycle there. The function $L_0(E,\epsilon)$ has integer-quantized one-sided slopes with respect to $\epsilon$ in this strip. The scalar term $I_d(\epsilon)$ is constant near $\epsilon=0$. Therefore, $f_E(\epsilon)=\widetilde L(E,\epsilon)$ also has integer-quantized slopes close to zero. The convexity and the two outer limits only allow slopes $-1$, $0$, and $+1$.

If Eq.~\eqref{methods:bound} were strict at a spectral energy, a slope $-1$ or $+1$ near zero would force the convex graph onto its corresponding outer supporting line, contradicting strictness. The only remaining local slope is zero. Since $\operatorname{Re}d(e^{\ii\theta})>0$ for $|a|<1$, the denominator has zero winding and $I_d(\epsilon)$ is constant near zero. The original Lyapunov exponent would thus be positive and locally constant under imaginary phase shifts. Avila's regularity criterion then makes the cocycle uniformly hyperbolic, which places $E$ outside the spectrum of the reciprocal Schr\"odinger operator. This contradiction forces equality in Eq.~\eqref{methods:bound}, and a subtraction of $\ln D$ yields Eq.~\eqref{eq:exactLE}.

\subsection{Imaginary gauge choice and boundary conditions}\label{methods:gauge}
For nonreciprocal hopping, the substitution $\psi_n=e^{-gn}\phi_n$ gives a reciprocal bulk equation. For a reciprocal localized state centered near $n_0$, this factor changes the decay rates of right and left tails to $\kappa_R=L_0(E)+g$ and $\kappa_L=L_0(E)-g$.
Both rates remain positive when $L_0(E)>|g|$. The weaker one vanishes at $L_0(E)=|g|$, yielding the threshold $R_g=De^{|g|}$. The finite-open-chain similarity transformation and the corresponding periodic-boundary relation are derived in Supplementary Note~4.

\subsection{Numerical comparisons}
\label{methods:numerics}
The ratios of consecutive Fibonacci numbers approximate the irrational frequency $\alpha=(\sqrt{5}-1)/2$. The system sizes and parameters are given in the figure captions. For the Lyapunov-exponent comparison in Fig.~\ref{fig:LEvalidation}, we first diagonalize the $N=377$ periodic chain with $\alpha=233/377$. We sort the eigenenergies by $L_{\mathrm{an}}$ and select $60$ entries approximately equally spaced in this order. For each selected energy, we iterate the original reciprocal transfer matrices for $30000$ steps using the irrational frequency $\alpha=(\sqrt{5}-1)/2$ and the single initial phase $\varphi=0$, without phase averaging. Starting from the vector $(1,0)^{\mathsf T}$, we normalize the propagated vector after each step and accumulate the logarithms of its Euclidean norm. Their average gives $L_{\mathrm{num}}$. The transfer-matrix iteration itself does not use the channel-radius formula, which is used only for the analytical comparison and the sampling order.

Right eigenstates are normalized by $\sum_n|\psi_n|^2=1$. {We evaluate the inverse participation ratio (IPR) and the finite-size spread indicator (fractal dimension) as}
\begin{equation}
\operatorname{IPR}=\sum_n|\psi_n|^4,\qquad D_2=-\frac{\ln\operatorname{IPR}}{\ln N}.
\label{methods:IPR}
\end{equation}
For increasingly large systems, an extended state tends towards $D_2=1$, while a localized state tends towards $D_2=0$. At finite sizes, $D_2$ is only a measure of how broadly a right eigenstate is distributed. Intermediate values do not demonstrate criticality. In particular, Fig.~\ref{fig:phase} presents the real parts of eigenenergies, so an apparent crossing in that plot must not be mistaken for the exact membership of a complex contour.

\section*{Acknowledgements}
L.Z. is supported by the National Natural Science Foundation of China (Grant No.~12275260), the Shandong Provincial Natural Science Foundation (Grant No.~ZR2026QB21), and the Fundamental Research Funds for the Central Universities (Grant No.~202364008). W.Z. is supported by the National Natural Science Foundation of China (Grants No.~12404499).

% Before submission, replace the commented placeholders below with the
% statements appropriate to the final manuscript.
%
% \section*{Data availability}
% State where the data supporting the findings are available.
%
% \section*{Code availability}
% State where the numerical code is available, if applicable.
%
% \section*{Author contributions}
% Describe the contribution of each author.
%
% \section*{Competing interests}
% State whether the authors have competing interests.

\end{document}